# Uncertainty quantification of two-phase flow in porous media via coupled-TgNN surrogate model

Jian Li[a], Dongxiao Zhang[b,c,*], Tianhao He[d], Qiang Zheng[c]

[a] China University of Petroleum (East China), Qingdao 266580, PR China;
[b] Southern University of Science and Technology, Shenzhen 518000, PR China;
[c] Peng Cheng Laboratory, Shenzhen 518000, PR China;
[d] Peking University, Beijing 100871, PR China;

## Abstract

Uncertainty quantification (UQ) of subsurface two-phase flow usually requires numerous executions of forward simulations under varying conditions. In this work, a novel coupled theory-guided neural network (TgNN) based surrogate model is built to facilitate computation efficiency under the premise of satisfactory accuracy. The core notion of this proposed method is to bridge two separate blocks on top of an overall network. They underlie the TgNN model in a coupled form, which reflects the coupling nature of pressure and water saturation in the two-phase flow equation. The TgNN model not only relies on labeled data, but also incorporates underlying scientific theory and experiential rules (e.g., governing equations, stochastic parameter fields, boundary and initial conditions, well conditions, and expert knowledge) as additional components into the loss function. The performance of the TgNN-based surrogate model for two-phase flow problems is tested by different numbers of labeled data and collocation points, as well as the existence of data noise. The proposed TgNN-based surrogate model offers an effective way to solve the coupled nonlinear two-phase flow problem and demonstrates good accuracy and strong robustness when compared with the purely data-driven surrogate model. By combining the accurate TgNN-based surrogate model with the Monte Carlo method, UQ tasks can be performed at a minimum cost to evaluate statistical quantities. Since the heterogeneity of the random fields strongly impacts the results of the surrogate model, corresponding variance and correlation length are added to the input of the neural network to maintain its predictive capacity. In addition, several more complicated scenarios are also considered, including dynamically changing well conditions and dynamically changing variance of random fields. The results show that the TgNN-based surrogate model achieves satisfactory accuracy, stability, and efficiency in UQ problems of subsurface two-phase flow.

## 1. Introduction

The prediction of two-phase flow in porous media is the result of a comprehensive process that usually includes geophysics, engineering management, and data interpretations (Lu and Zhang 2006, Liang et al. 2018, Liu et al. 2019). Due to the inaccessibility and complexity of reservoirs, sufficient observational data are not obtained, leading to uncertainties in the heterogeneous parameter fields. As a consequence, the prediction of two-phase flow in porous media is essentially a process that is subject to a variety of uncertainties. The Monte Carlo method is commonly used to solve uncertainty quantification (UQ) problems, and reservoir simulation is the most typical means of forward simulation in oilfields (Maldonado-Cruz and Pyrcz 2022). As time has passed, reservoir models have become large and complex, and each simulation requires solving partial differential equations. However, many studies demand hundreds of forward runs of the simulator to provide meaningful coverage (Bao et al. 2020, Mo et al. 2020), which results in significant computational consumption for reservoir numerical simulation. Furthermore, when conditions change (e.g., boundary conditions, initial conditions, well control conditions, model size, prediction time, etc.), the time-consuming iterative solving process needs to restart again. As an alternative to repetitive forward simulations, surrogate models are used extensively across different scientific disciplines to approximate the results of numerical simulations and reduce computational cost (Zhu and Zabaras 2018, Jin et al. 2020, Tang et al. 2021). Therefore, the purpose of this work is to build a concise, accurate, and efficient surrogate model of two-phase flow in porous media.

Surrogate models, also known as proxy models, can obtain mapping relationships between inputs and outputs through black-box calculations to improve the efficiency of positive prediction with minimal accuracy cost. The prominent ways of constructing surrogate models in the petroleum field mainly include polynomial chaos expansion and Gaussian processes (Alkhatib and King 2014, Babaei et al. 2015, Li et al. 2015). Although these methods have achieved good results in practical applications, they are limited to scenarios with relatively low-dimensional inputs, as they both suffer from the curse of dimensionality. As the simulation dimension increases, the limitations of these methods become increasingly evident. In truncated polynomial chaos expansion, the number of required basis functions increases with the number of orders and dimensions, which leads to a substantial increase in computational effort. A common selection of the total number of basis functions is equal to $(n+k)!/(n!k!)$ (Li et al. 2014), where n is the number of dimensions; and k is the order of expansion. Meanwhile, due to its non-parametric learning property, the Gaussian process method will become increasingly complex as the amount and dimensionality of data increase, and computation on full rank kernel matrices is highly demanding.

With the improvement in hardware computing power and rapid growth of data volume, deep neural network (DNN) applications in the construction of surrogate models have advanced dramatically (Seong et al. 2020, Shen et al. 2022, Xiao et al.

2022). The three main superiorities of DNNs are as follows: 1) DNNs can approximate any combination of logic functions by the addition of nonlinear activation functions (Yarotsky 2017); 2) DNNs have an efficient operation speed due to their special computation scheme. A forward calculation only takes a few seconds, and sometimes even less (Miller et al. 2017, Zhong et al. 2022). 3) In traditional approximation methods, the prediction error generally comes from the data dimension. The data dimension is usually in the exponential position of the error expression, which means that the higher is the dimension, the larger is the error. DNN, however, essentially does the function approximation. In the approximation process, the error is not related to the dimension, and thus it can be effective at dealing with high-dimensional problems (Han et al. 2018). The above-mentioned advantages give this approach unique properties to better handle high-dimensional problems efficiently.

This NN-based framework has been broadly used to build surrogate models due to its fast forward computation and ability to learn complex associations between inputs and outputs. Ma et al. (2022) proposed a spatial-temporal convolution recurrent neural network surrogate model for history-matching problems. The model was based on a densely connected convolutional neural network (CNN) model and a stacked multilayer long short-term memory (LSTM) structure. Zhang et al. (2021) introduced a fully convolutional encoder/decoder network based on DenseNet to establish a surrogate model for predicting water saturation fields, and they innovatively considered different sizes of permeability fields as input features. The above studies proved that DNNs can work well as a surrogate model with high efficiency and accuracy. However, the previous models are trained in a data-driven scheme, which requires a large amount of field data to train the network parameters. When the permeability field or the selected area changes, the training data need to be re-collected for retraining. On the one hand, the data are hard to obtain and very expensive in practice (Chen and Zhang 2020, Valdez et al. 2020); on the other hand, labeled data obtained from the real world are usually sparse and incomplete, and therefore may not be suitable for DNN models built in ideal environments. In addition, data-driven models lack certain generalization, such as poor predictive performance at extrapolated time points or the presence of points that violate expert knowledge. Therefore, some supporting physical information, including scientific theories, governing equations, and expert knowledge, should be incorporated to adequately address these issues.

The theory-guided machine learning framework has recently emerged under such conditions (Karpatne et al. 2017, Raissi et al. 2018). Li et al. (2022) developed a theory-guided neural network (TgNN) as a prediction model for two-phase flow, in which two independent neural networks were built in the TgNN framework, with one approximating pressure and the other approximating saturation, respectively. After a two-stage strategy for the training process, scientific theories could be well incorporated into neural networks. However, it was a predictive model, rather than a surrogate model, and uncertainty in the permeability field and other inputs were not included in training. Wang et al. (2021a) developed the TgNN as a surrogate model for dynamic subsurface single-phase flow, in which only fluid pressure is considered. Later, they extended the single-phase flow surrogate model to a two-phase flow (Wang et al. 2021b). In this work, they proposed a theory-guided convolutional neural network (TgCNN) for UQ problems of two-phase flow in porous media. They used two dependent CNNs to simultaneously approximate the two principal variables of the considered problem. The two networks are mutually conditioned in the training process by the discretized governing equation. Since CNNs cannot use auto-differentiation, the approach of TgCNN to solve partial differential equations (PDEs) is generally to apply a finite difference form. This method discards the auto-differentiation, of which the neural network is effective, resulting in relatively lower computational speed. Furthermore, most of the current deep learning training schemes for coupled PDE calculations do not provide a detailed comparison of the influence of network hyperparameters, resulting in a rigid and redundant network structure. This subsequently leads to the inability of the network to change the hyperparameters differently according to the parameters with the two different regression methods, i.e., pressure and water saturation.

To render the model more consistent with physical laws, a surrogate model constructed by the coupled theory-guided neural network (TgNN) is proposed as a predictive process for oil-water phase flow in this work. The coupled TgNN consists of a coupled block and two independent blocks approximating pressure and water saturation, respectively. The model is driven by not only labeled data, but also scientific theories, including governing equations, stochastic parameter fields, boundary and initial conditions, well conditions, and expert knowledge. The accuracy and robustness of the surrogate model are tested, and then UQ tasks are performed by combining the model with the MC method. Since the variation in random fields' heterogeneity exacerbates the difficulty of the surrogate model in terms of accuracy and robustness, corresponding variance and correlation length are added to the input of the neural network to maintain its predictive capacity. In addition, several more complicated scenarios are tested, including the existence of data noise, dynamically changing well conditions, and dynamically changing variance of parameter fields. The results show that the TgNN-based surrogate model possesses satisfactory accuracy, stability, and efficiency in UQ problems of subsurface two-phase flow.

The remainder of this paper is organized as follows. The **Methodology** section provides the architecture and mechanism of the TgNN-based surrogate model, including screening and constructing neural network models. The **Case Studies** section investigates the accuracy and generalizability of the TgNN-based surrogate model. Then, we test the performance of the surrogate model constructed for a two-phase subsurface flow problem in several complicated scenarios. The **Discussion and Conclusion** section summarizes the main contents of this paper and identifies potential directions for future work.

## 2. Methodology

In this work, subsurface flow of the oil/water phase in two-dimensional heterogeneous porous media is considered.

## 2.1 Model structure

The DNN model is a powerful function approximator, which aims to establish a mapping relationship between the input layer (X) and the output layer (Y). The residual, known as data mismatch between labeled and generated data, can be represented in mean square error (MSE) form, which can be expressed as follows:

$$\text{MSE}_{\text{data}}(\theta) = \frac{1}{N}\sum_{i=1}^{N}\left|N(x_i;\theta) - y_i\right|^2, \tag{1}$$

The TgNN framework for oil-water two-phase flow problems has been proposed in our previous work (Li et al. 2022), in which two independent neural networks are built in TgNN for oil-water phase flow problems, with one approximating pressure and the other approximating saturation. The predictions of TgNN possess sufficient accuracy. However, the independence between the two neural networks leads to a long training time, which is prohibitive for surrogate models. Therefore, in order to improve accuracy and simultaneously shorten the training time, here we firstly perform some screening and construction of the networks.

To reflect the calculation of the coupling nature of pressure and water saturation, and distinguish the difference between the regression methods of the two parameters in the actual calculation, the following four kinds of coupled network models are constructed: 1) one integrated network with two outputs; 2) two independent networks with two outputs; 3) coupled networks that merge first and then separate; and 4) coupled networks that separate first and then merge. A schematic diagram of the network with different styles is shown in **Figure 1**.

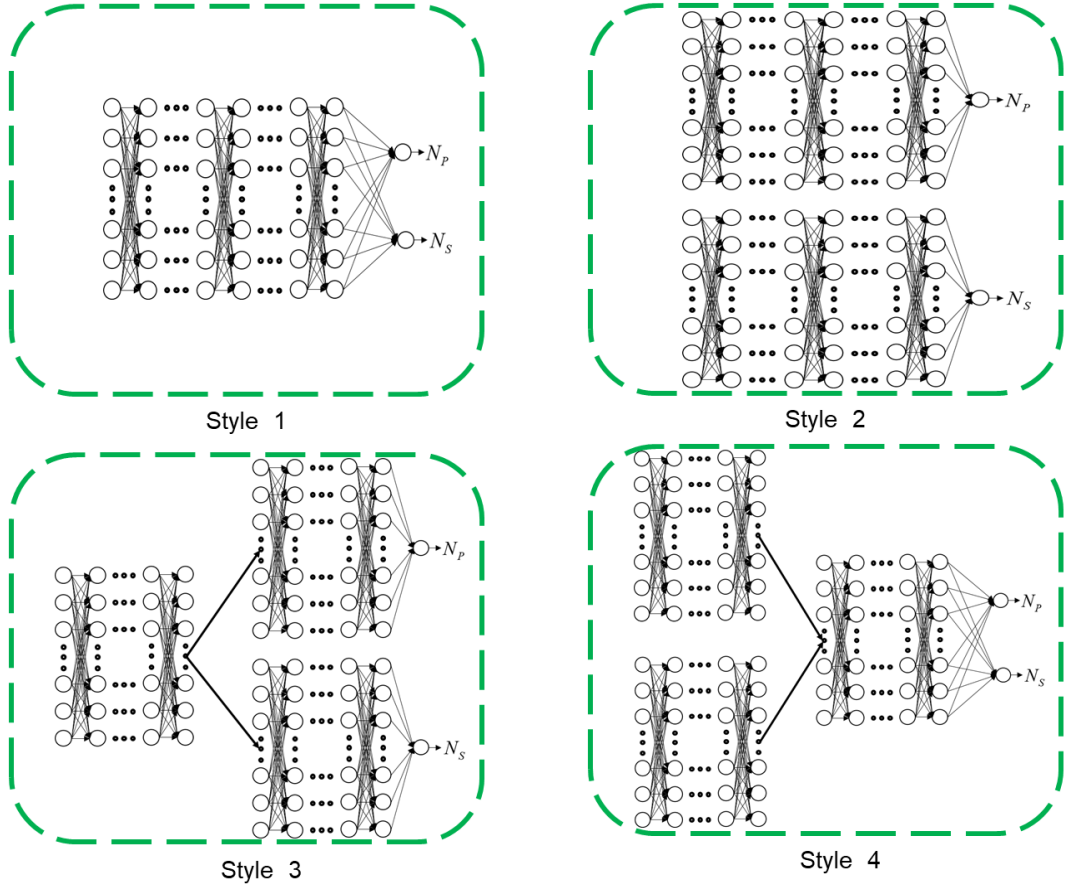

**Figure 1.** Schematic diagram of four kinds of network construction.

To reduce extraneous interference, the parameters of neural networks are unified, in which the number of layers is eight and the number of neurons in each layer is 100. The input layer consists of spatial-temporal points, as well as permeability field parameters, and the output layer contains pressure and water saturation. There are 30 permeability fields being used as labeled data, and 100 other ones are used as test data to evaluate the generalizability of the model. The network models are tested in a purely data-driven scheme, with 500 training epochs. Relative $L_2$ error and $R^2$ score are employed as metrics to quantitatively evaluate the performance of the four network models, which are presented as follows:

$$L_2\left(u_{pre}, u_{true}\right) = \frac{\|u_{pre} - u_{true}\|}{\|u_{true}\|}$$

$$R^2 = 1 - \frac{\sum_{n=1}^{N}\left(u_{pre,n} - u_{true,n}\right)^2}{\sum_{n=1}^{N}\left(u_{true,n} - \overline{u_{true}}\right)^2}, \tag{2}$$

where $\|\cdot\|$ denotes the standard Euclidean norm; $u_{true}$ is the reference value obtained from the numerical simulation; $u_{pre}$ is the predicted value obtained from the neural networks; and $n$ represents the index of predictions.

We use $\overline{L_2}$ and $\overline{R^2}$ to represent the mean relative $L_2$ error and the mean $R^2$ score of pressure and water saturation, respectively. The evaluations of pressure and water saturation in 100 test data are shown in **Table 1**. The network models are screened according to their accuracy and time consumption.

**Table 1.** $\overline{L_2}$ and $\overline{R^2}$ of pressure and water saturation in 100 test data.

| Style | $\overline{L_2}$ | $\overline{R^2}$ | Time (min) |
|---|---|---|---|
| 1 | 5.27746e-02 | 7.88171e-01 | 4:03 |
| 2 | 5.38114e-02 | 8.84395e-01 | 9:11 |
| 3 | 4.88945e-02 | 8.42185e-01 | 3:18 |
| 4 | 5.12429e-02 | 7.52079e-01 | 5:39 |

It can be seen that **Style 3** performs the best in both accuracy and training speed. After leveraging the control variable method for hyperparameters, the optimal network structure is obtained. As shown in the schematic structure of TgNN in **Figure 2**, four hidden layers with 100 neurons in each layer are used as the calculation of the coupling part, four hidden layers with 100 neurons in each layer are used in $NN_P$ to regress the pressure, and four hidden layers with 50 neurons in each layer are used in $NN_S$ to regress the water saturation.

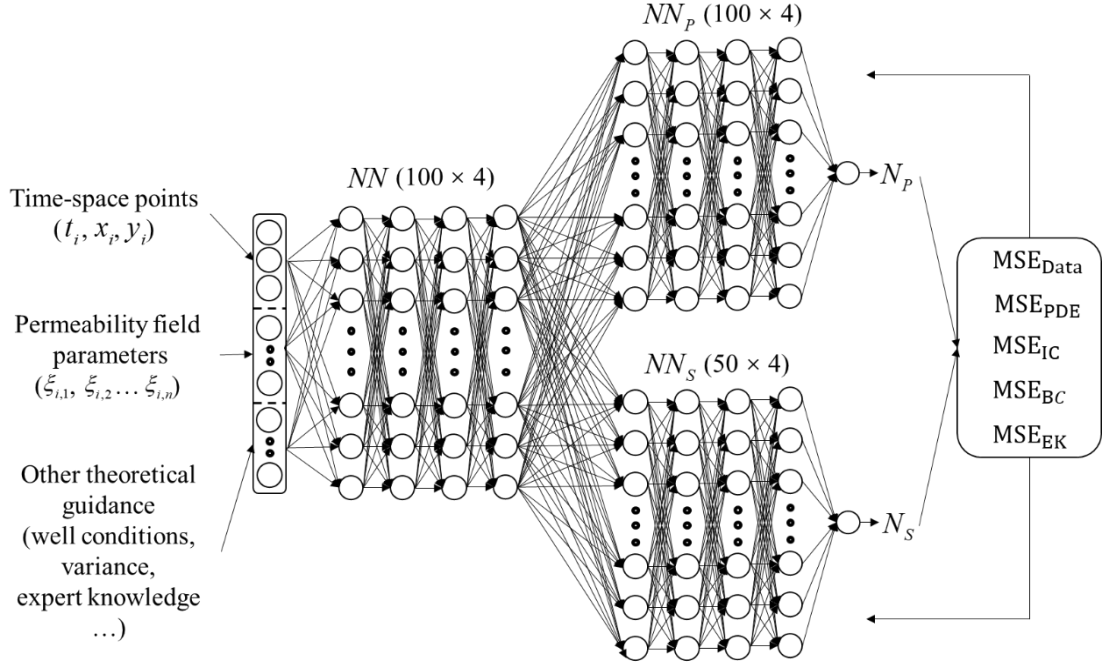

**Figure 2.** Schematic diagram of the TgNN-based surrogate model.

## 2.2 Governing equations

The governing equations of subsurface flow for oil/water phase in 2D heterogeneous porous media are shown as follows:

$$\frac{\partial}{\partial x}\left(\rho_w \frac{kk_{rw}}{\mu_w}\frac{\partial P_w}{\partial x}\right) + \frac{\partial}{\partial y}\left(\rho_w \frac{kk_{rw}}{\mu_w}\frac{\partial P_w}{\partial y}\right) + Q_w = \frac{\partial}{\partial t}(\phi \rho_w S_w)$$
$$\frac{\partial}{\partial x}\left(\rho_o \frac{kk_{ro}}{\mu_o}\frac{\partial P_o}{\partial x}\right) + \frac{\partial}{\partial y}\left(\rho_o \frac{kk_{ro}}{\mu_o}\frac{\partial P_o}{\partial y}\right) + Q_o = \frac{\partial}{\partial t}(\phi \rho_o S_o) \;, \quad (3)$$
$$P_C = P_o - P_w$$
$$S_o + S_w = 1$$

where $P_o$ and $P_w$ denote the pressure of oil and water phase, respectively; $S_o$ and

$S_w$ denote the saturation of oil and water phase, respectively; $P_C$ represents the capillary pressure of the reservoir; $\rho_o$ and $\rho_w$ represent oil and water density, respectively; $k$ is permeability; $k_{ro}$ and $k_{rw}$ are the relative permeability of oil and water, respectively; $\mu_o$ and $\mu_w$ are oil and water viscosity, respectively; $\phi$ is the porosity of the reservoir rock; $Q_o$ and $Q_w$ represent the source/sink term of the oil and water phase, respectively, which only exists in the injection well or the production well.

$$Q_l = 2\pi h k k_{rl} \rho_l \times \frac{P_l - P_{wf}}{\mu_l \ln(\frac{r_e}{r_w})} \quad (l = w, o) \quad , \tag{4}$$

where $r_w$ denotes the wellbore radius; and $r_e$ denotes the equivalent radius of the well block. For an isotropic system, $r_e$ is simplified (Peaceman 1978) as follows:

$$r_e = 0.14(\Delta x^2 + \Delta y^2)^{1/2} \tag{5}$$

Therefore, the residual of the governing equation in the water phase and the oil phase can be added into the loss function of the neural network as a regularized term as follows:

$$\begin{aligned} f_l &= \frac{\rho_l}{\mu_l} \frac{\partial (kk_{rl})}{\partial x} \frac{\partial N_P(t,x,y;\theta)}{\partial x} + \frac{\rho_l kk_{rl}}{\mu_l} \frac{\partial^2 N_P(t,x,y;\theta)}{\partial x} \\ &+ \frac{\rho_l}{\mu_l} \frac{\partial (kk_{rl})}{\partial y} \frac{\partial N_P(t,x,y;\theta)}{\partial y} + \frac{\rho_l kk_{rl}}{\mu_l} \frac{\partial^2 N_P(t,x,y;\theta)}{\partial y} + 2\pi h k k_{rl} \rho_l \times \frac{P_l - P_{wf}}{\mu_l \ln(\frac{r_e}{r_w})}, \\ &- \phi \rho_l \left( C_\phi N_S(t,x,y;\theta_S) \frac{\partial N_P(t,x,y;\theta)}{\partial t} + \frac{\partial N_S(t,x,y;\theta)}{\partial t} \right) \quad l = w, o \end{aligned} \tag{6}$$

where $C_\phi$ is the compressibility of the reservoir rock. By adopting automatic differentiation, it is efficient to calculate the partial derivative of the outputted $N_P(t,x,y;\theta)$ and $N_S(t,x,y;\theta)$ by the neural network with respect to the input.

Therefore, the partial derivatives of $N_P(t,x,y;\theta)$ and $N_S(t,x,y;\theta)$ with respect to time and space can be rapidly computed, and the partial derivatives of $k_{rl}$ with respect to space can be transformed with water saturation by utilizing the relative permeability curve. Additional details about the derivation of the equations can be found in our previous work (Li et al. 2022).

To constrain the governing equation, $f_l$ should be added into the loss function as follows:

$$MSE_{PDE}(\theta) = \frac{1}{N_{f_l}} \sum_{i=1}^{N_{f_l}} \left| f_l(t_f^i, x_f^i, y_f^i) \right|^2 \quad (l = w, o), \tag{7}$$

where $(t_f^i, x_f^i, y_f^i)$ denotes the coordinates of collocation points; and $N_{f_l}$ denotes the number of collocation points for $f_l$.

Furthermore, the residual of boundary conditions and initial conditions can also be calculated in the same way. Therefore, the total loss function of the TgNN-based surrogate model for the two-phase flow problem can be formulated as follows:

$$L(\theta) = \lambda_{data}MSE_{data}(\theta) + \lambda_{PDE}MSE_{PDE}(\theta) + \lambda_{IC}MSE_{IC}(\theta) + \lambda_{BC}MSE_{BC}(\theta) + \lambda_{EK}MSE_{EK}(\theta), \tag{8}$$

where $MSE_{IC}$, $MSE_{BC}$, and $MSE_{EK}$ are the mean square error of the loss functions of the initial condition term, the boundary condition term, and the expert knowledge term, respectively; and $\lambda$ are weights to adjust the relative importance of corresponding terms in the loss function.

The learning rate determines the update step of parameters in the training process, which is initially set as 0.001. In order to converge to the global optimal value efficiently with high possibility, the learning rate continuously decreases exponentially with the number of training rounds. Consequently, a small learning rate will be used for restraining the output in a small area near the optimal value, or at least oscillating within an optimum range.

## 2.3 Parameterization of permeability fields with Karhunen-Loève expansion

In subsurface flow problems, permeability fields are usually heterogeneous, which can be dealt with as stochastic fields. The Karhunen-Loève expansion (KLE) in (Chang and Zhang 2014) can be employed to parameterize stochastic fields with specific distribution and covariance.

For a stochastic field of permeability $Z(x,y,\tau) = \ln k(x,y,\tau)$, where $(x,y) \in D$ (physical domain) and $\tau \in \Theta$ (a probability space), the random field can be expressed with a finite number of terms as follows:

$$Z(x,y) \approx \overline{Z(x,y)} + \sum_{i=1}^{n} \sqrt{\lambda_i} f_i(x,y) \xi_i(\tau), \qquad (9)$$

where $n$ denotes the number of truncated terms in the KLE; and $\lambda_i$ and $f_i(x,y)$ are the eigenvalue and eigenfunction of the covariance, respectively. Additional details about the method can be found in (Zhang and Lu 2004).

The random field can then be represented by:

$$k(x,y) = \exp\left(\overline{k(x,y)} + \sum_{i=1}^{n} \sqrt{\lambda_i} f_i(x,y) \xi_i(\tau)\right) \qquad (10)$$

The above expression of permeability fields will be utilized for calculating the differentiation of permeability fields with respect to the coordinates in the residual of governing equations, as shown in equation (6).

# 3. Case Study

In this section, the performance of the TgNN-based surrogate model for oil-water phase flow in two-dimensional heterogeneous porous media is tested. Several scenarios, including different training numbers of labeled data and collocation points, the existence of data noise, the heterogeneity of permeability fields with different conditions, and changes in well condition and variance are investigated.

We consider a square domain with a size 510 m × 510 m, which is divided into 51 × 51 grids. The outer boundary conditions are assigned as four no-flow boundaries, and all of the points on the outer boundaries should obey the Newman boundary condition. There is an injection well and a production well located at two grids with coordinates as (1, 1) and (51, 51), respectively. The porosity of this field is 20%, and the compressibility of reservoir rock is $3 \times 10^{-6}$ /kPa. The initial conditions are $P_{t=0}=100$ bar and $S_{w_{t=0}} = 0.2$. The viscosity of water and oil is 1 mPa·s and 5 mPa·s, respectively. To simplify the problem, the capillary pressure is 0, which means $P_o=P_w$. The total simulation time is 1800 d, and each time-step is 36 d, resulting in 50 time-steps. The labeled dataset is obtained from UNCONG software (Liu et al. 2022) simulation results.

The coupled network uses a sequential training strategy, which means that the labeled data are trained first at each training epoch, and on the basis of that, the constraints of other scientific theories are trained by adding collocation points. The Softplus activation function is employed, and the Adam algorithm optimizer is used in the training process to tail parameters.

## 3.1 Surrogate model for two-phase flow in porous media

In this case, both the injection wells and production wells are operated to follow bottom hole pressure (BHP) controls. Unless otherwise stated, the bottom hole pressure of the injection well is 150 bar and the production well is 80 bar at all time-steps. The correlation length of this physical domain is 204 m, and the mean and variance of the log-transformed permeability field are set as $\overline{\ln k(x,y)} = 6.5$ and $\sigma^2_{\ln k(x,y)} = 0.5$, respectively. We maintain 20 terms in KLE (i.e., $n = 20$) to preserve 80% energy of the random permeability field.

### 3.1.1 Accuracy of the prediction and UQ tasks

Unless otherwise stated, 500 permeability fields are generated to participate in the training stage, with 50 labeled datasets for data mismatch and 450 non-labeled datasets for theoretically guided constraints. Then, 5,000 permeability fields are generated for testing. The input layer of the TgNN-based surrogate model is composed of time-space points ($t_i$, $x_i$, $y_i$) and 20 parameters ($\xi_{i,1}$ ... $\xi_{i,20}$) representing the permeability field, where $i$ represents the index of permeability field. Therefore, the dimension of input is 23. We randomly selected 2,000,000 observation data from 50 permeability fields as the labeled training set. In addition to data mismatch, the theory-guided information, i.e., governing equations, boundary and initial conditions, and expert knowledge, is also employed in the training process. In the calculation process of $MSE_{PDE}$, 3,000,000 data are randomly selected from 500 permeability fields as collocation points. In the calculation of $MSE_{IC}$, 15,000 data are randomly extracted at the initial moment in 500 permeability fields as collocation points; in the calculation of $MSE_{BC}$, 25,000 data are randomly chosen at the boundary condition in 500 permeability fields as collocation points. The network training is performed on an NVIDIA GeForce RTX 2060 Graphics Processing Unit (GPU) card, and 1,500 training epochs are carried out to optimize the loss function, which takes approximately 3.04 h (10,953 s).

**Figure 3** shows the correlation line between the predictions of the two models with two points taken randomly from the 200 permeability fields in the test set. The results show that the TgNN-based surrogate model can match well with the reference data, and they almost lie on a straight line at an angle of 45°. **Figure 4** presents the comparison of pressure fields and saturation fields among the three stochastic permeability realizations at time-step 30 between the TgNN-based surrogate model and the reference. As can be seen from the figure, the TgNN-based surrogate model is effective when faced with different permeability fields, and the errors are within acceptable limits. To better represent the overall comparison of the TgNN-based surrogate model and the purely data-driven surrogate model, we select 100 permeability fields in the test set to calculate the relative $L_2$ error and $R^2$ score, as shown in **Figure 5.** The results demonstrate that the TgNN-based surrogate model makes better predictions in terms of both relative $L_2$ error and $R^2$ score.

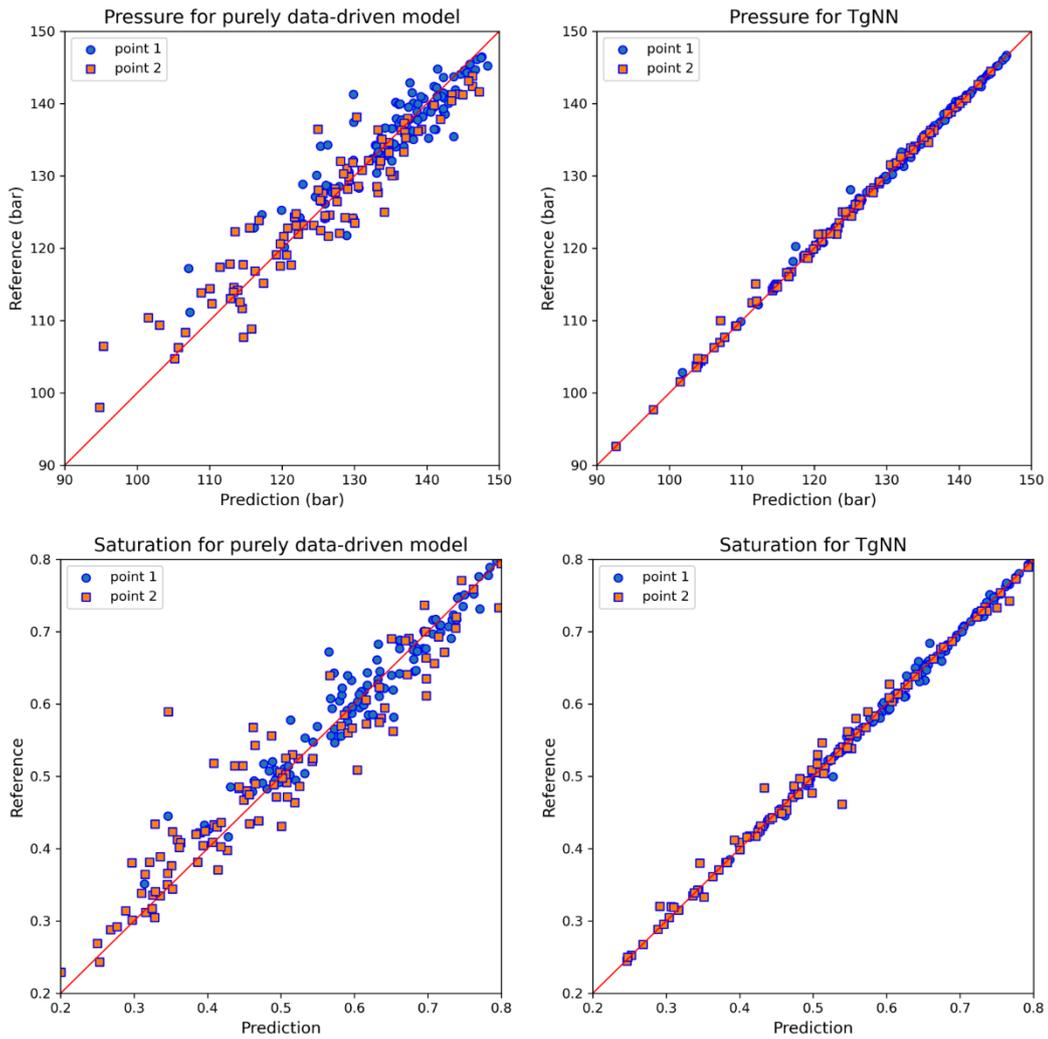

**Figure 3.** The correlation line between the predictions of the two models with two points taken randomly from the 200 permeability fields in the test set.

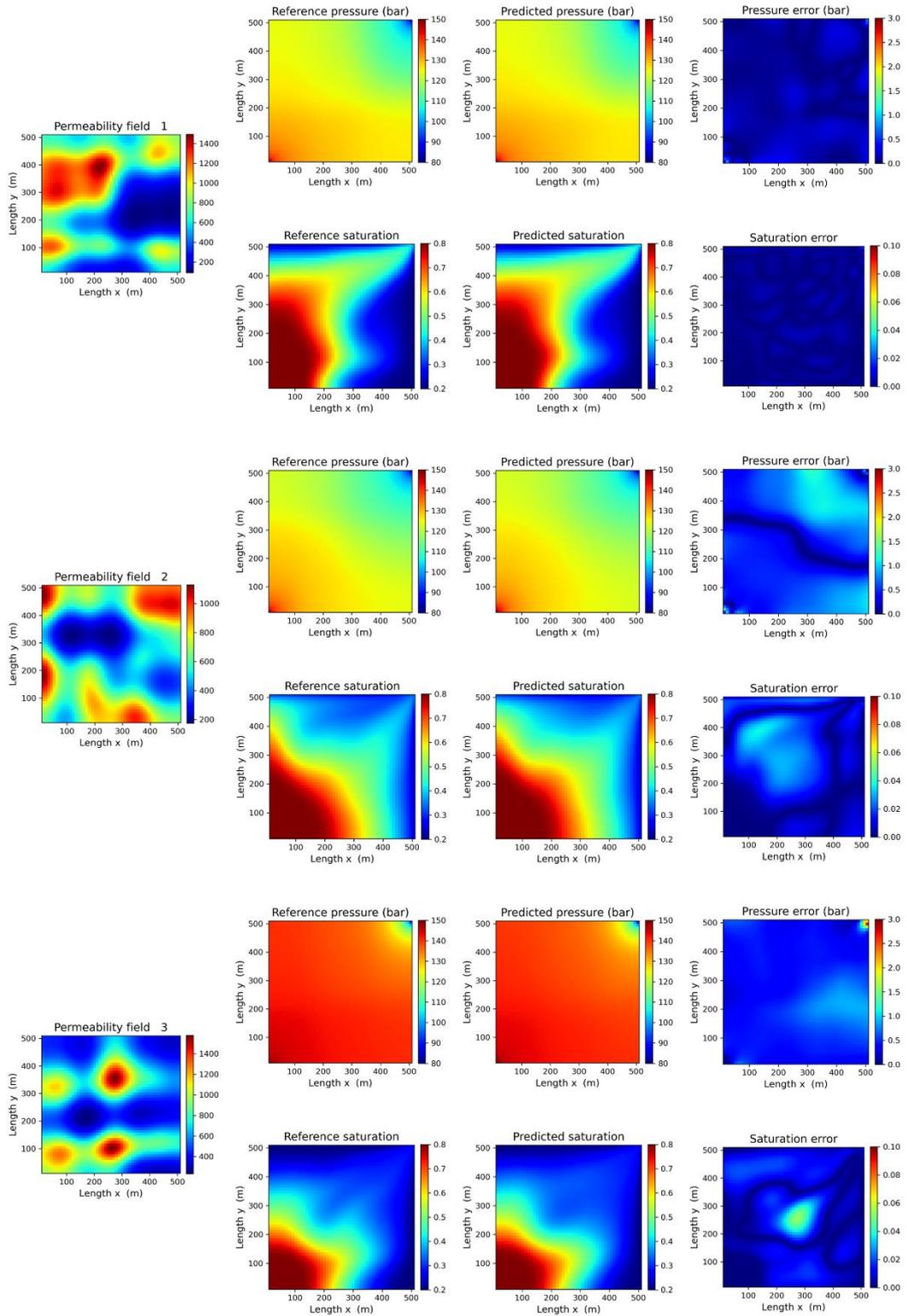

**Figure 4.** Comparison of pressure fields and saturation fields among the three stochastic permeability realizations at time-step 30 between the TgNN-based surrogate model and the reference.

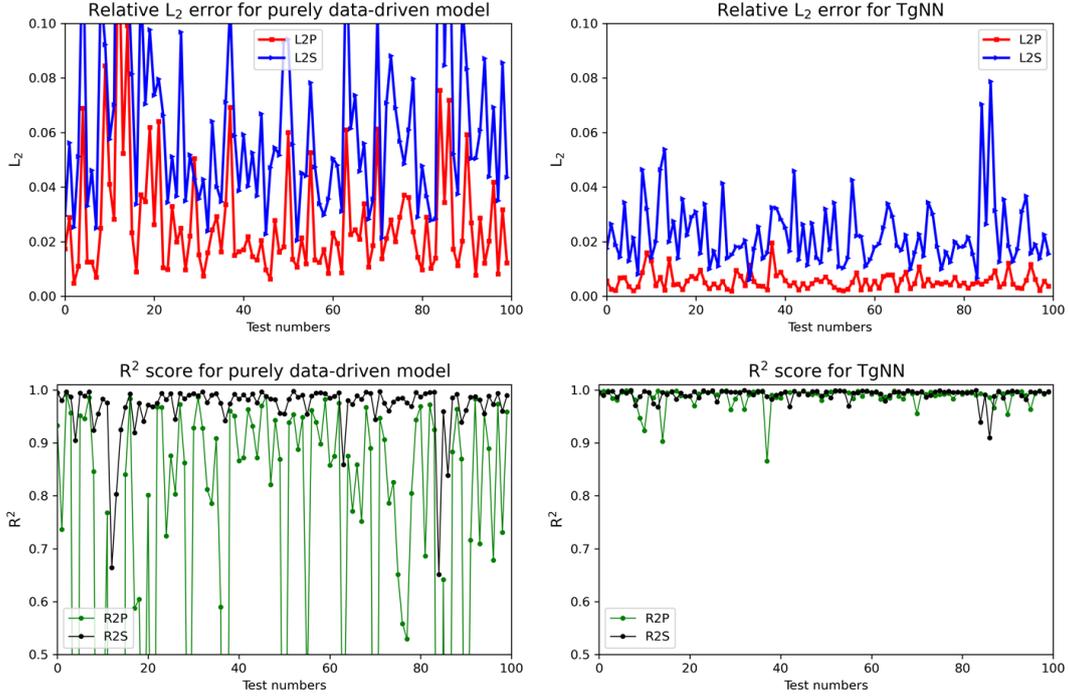

**Figure 5.** Relative $L_2$ error and $R^2$ score acquired by the TgNN-based surrogate model and the purely data-driven surrogate model from 100 permeability fields in the test set.

By combining the accurate TgNN-based surrogate model with the MC method, UQ tasks can be performed at a minimum cost to evaluate statistical quantities of two-phase flow in porous media. The reference data are obtained from 5,000 realizations solved with UNCONG software, and the same number of predicted data are also available by performing forward operations of the TgNN-based surrogate model. The mean, variance, and the probability density function (PDF) of the pressure and water saturation from the reference and predicted data are used as a test for UQ tasks. **Figure 6** shows a comparison of the mean and variance of the surrogate outputs and the reference at time-step 30. **Figure 7** presents the estimated PDF of two fixed points in 5,000 realizations. The results indicate that both the pressure and saturation from the outputs of the TgNN-based surrogate model can match well with the reference values in terms of mean, variance, and PDF. This indicates that the TgNN-based surrogate model is able to very accurately capture the relationship between input and output. It is also worth noting that the predictive efficiency of the surrogate model is even more impressive. The forward calculation time for the TgNN-based surrogate model is approximately 0.5 h (1,965 s), and the training time of the model is 3.2 h. However, it takes 6.2 h for the simulator to execute 5,000 realizations, which is much longer than the total time of the surrogate model (3.7 h). This indicates that, once trained, the TgNN-based surrogate model can perform forward calculations more efficiently.

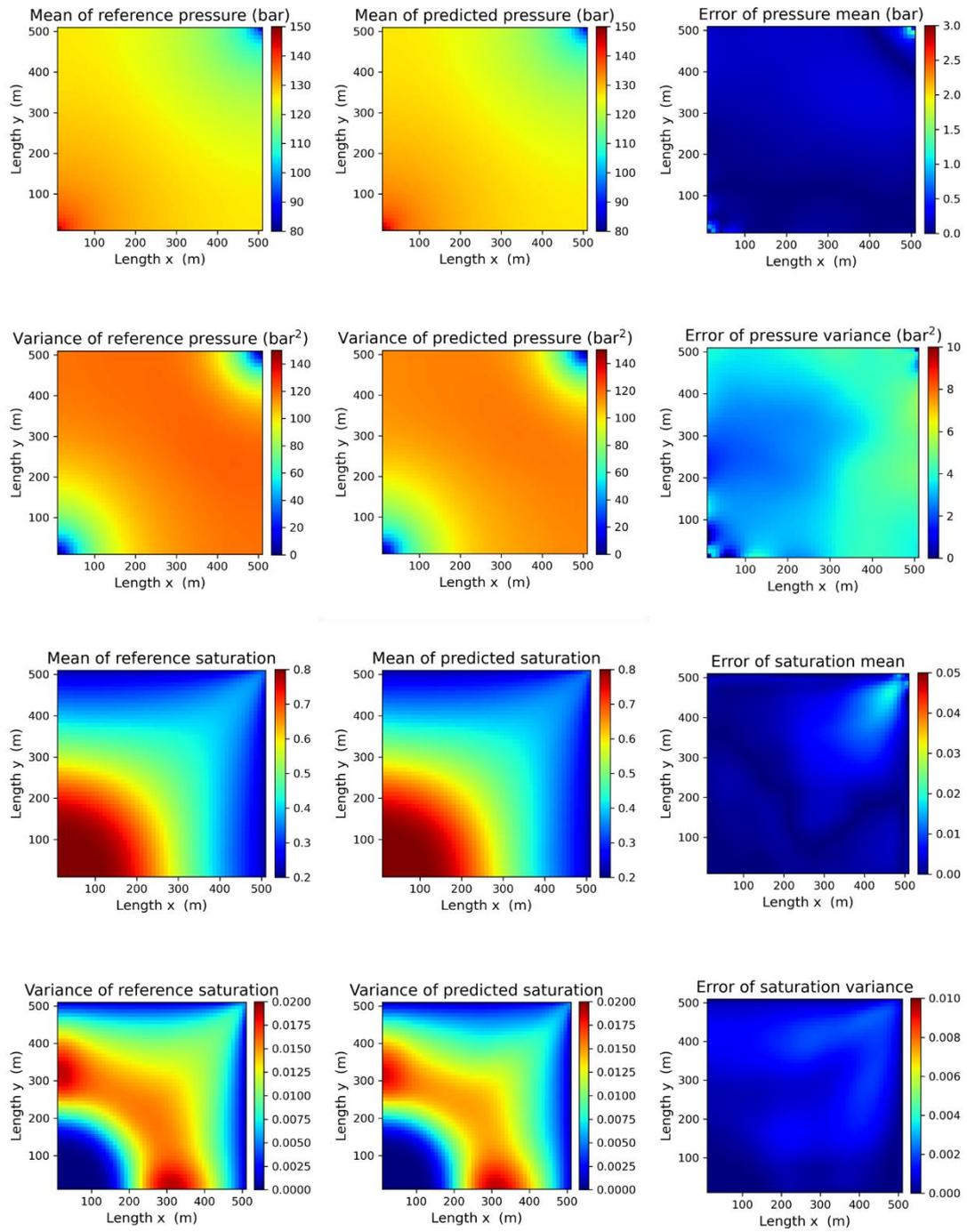

**Figure 6.** The mean and variance of pressure fields and saturation fields among 5,000 stochastic permeability realizations at time-step 30 between the TgNN-based surrogate model and the reference.

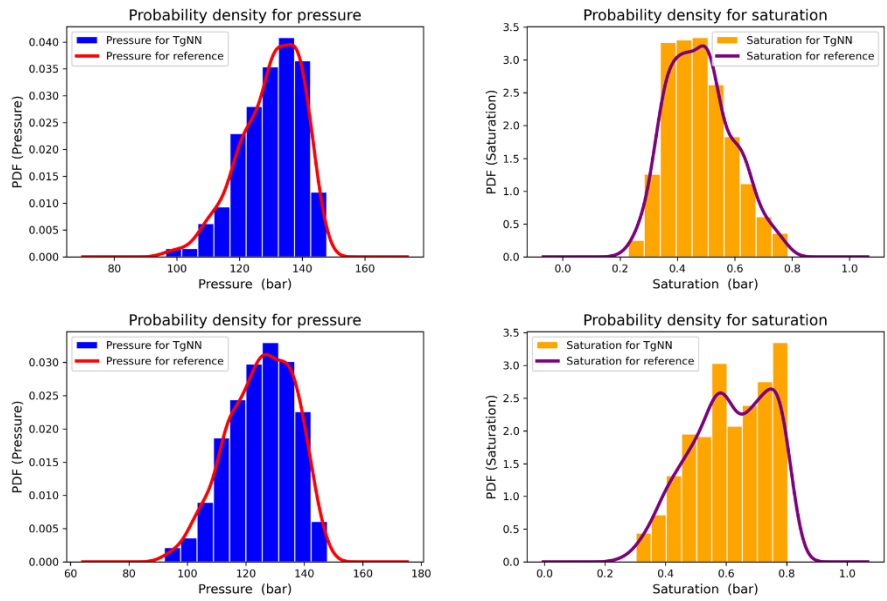

**Figure 7.** The estimated PDF of two fixed points (Point 1: x = 200 m, y = 200 m, t = 720 d; Point 2: x = 300 m, y = 400 m, t = 180 d) among 5,000 realizations.

**3.1.2 The impact of labeled data and collocation points**

In subsection 3.1.1, 2,000,000 labeled data are randomly taken from 50 labeled permeability fields, and 3,000,000 collocation points are randomly taken from 500 realizations. In this section, different amounts of labeled data and collocation points are used to examine the accuracy and time consumption of the TgNN-based surrogate model. Firstly, the number and location of collocation points are fixed (i.e., 3,000,000), and the influence of the number of labeled data is researched in three cases. The number of collocation points is then changed with the fixed setup of labeled data (i.e., 2,000,000). The result is presented in **Table 2**. As can be seen, with the increase of labeled data and collocation points, the TgNN-based surrogate model can provide more accurate predictions. It is also worth noting that the impact of the two types of data on the network is different.

From the perspective of model accuracy, labeled data do play a role in improving the effect as their number increases. However, the effect is relatively small, and the worst parameter $R^2_{pressure}$ is 0.965977 when the number of labeled data is minimal, which indicates that the pressure data have been fitted well. In contrast, the effect of collocation points on the accuracy of the surrogate model is particularly significant. Specifically, as the number of collocation points increases, significantly better prediction results can be seen by the improvement of each error metric.

From the perspective of time consumption, the influence of the amount of labeled data is limited. The three cases regarding the amount of labeled data show that the time consumption is almost within the same order of magnitude. In contrast, the number of collocation points is decisive for the training time. The automatic differentiation of partial differential terms in PDE and boundary conditions consumes a large amount of computational time (Wang et al. 2020), leading to an increase in training time as the number of collocation points increases. As a consequence, it is crucial to make a tradeoff between accuracy and computational cost by selecting an optimal number of collocation points.

After the optimization of accuracy and training efficiency, 2,000,000 labeled data and 3,000,000 collocation points are chosen to perform the prediction of 5,000 permeability fields. The results of relative error and $R^2$ score for pressure and water saturation at each time-step are shown in **Figure 8**. It can be seen that both pressure and the water saturation exhibit a good fitting performance with small error distribution and high stability.

**Table 2.** Accuracy and time consumption of the TgNN-based surrogate model using different numbers of labeled data and collocation points.

| Style | Numbers | $L_{2-pressure}$ | $L_{2-saturation}$ | $R^2_{pressure}$ | $R^2_{saturation}$ | Time (s) |
|---|---|---|---|---|---|---|
| Labeled data | 300,000 | 0.008109 | 0.041632 | 0.965977 | 0.987898 | 10,565 |
| | 1,000,000 | 0.007266 | 0.045289 | 0.971995 | 0.985496 | 10,693 |
| | 2,000,000 | 0.005505 | 0.034194 | 0.987296 | 0.991413 | 10,953 |
| **Collocation points** | 300,000 | 0.021097 | 0.055007 | 0.755229 | 0.977946 | 3,659 |
| | 3,000,000 | 0.005505 | 0.034194 | 0.987296 | 0.991413 | 10,953 |
| | 6,000,000 | 0.004991 | 0.031527 | 0.990320 | 0.995211 | 19,288 |

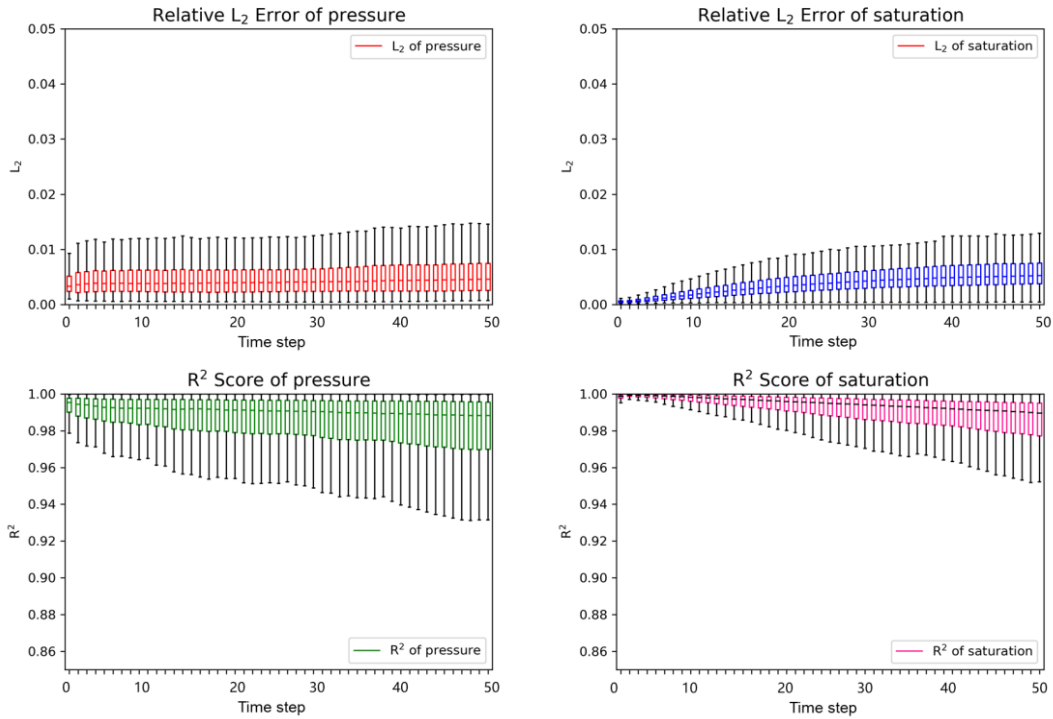

**Figure 8.** Relative $L_2$ error and $R^2$ score of pressure and water saturation at each time-step among 5,000 stochastic permeability realizations.

### 3.1.3 The impact of noise data

Noise is a ubiquitous problem in actual production. In actual oilfield production, labeled production data often fluctuate due to changes in instrumentation, human operations, and production strategies. In numerical calculations, these fluctuations are treated as measurement noise. The presence of noise affects the quality of the labeled data in the data mismatch of the model during training and further impacts the accuracy of the predictions. Consequently, anti-noise tests are necessary for surrogate models in terms of practical applications.

In this case, a 15% noise level is added to the full-time period of pressure and saturation data calculated by the labeled permeability fields. These noisy data are applied to the training process of the network through data mismatching which, in turn, affects the output accuracy. The method of adding noise can be found in our previous work (Li et al. 2022).

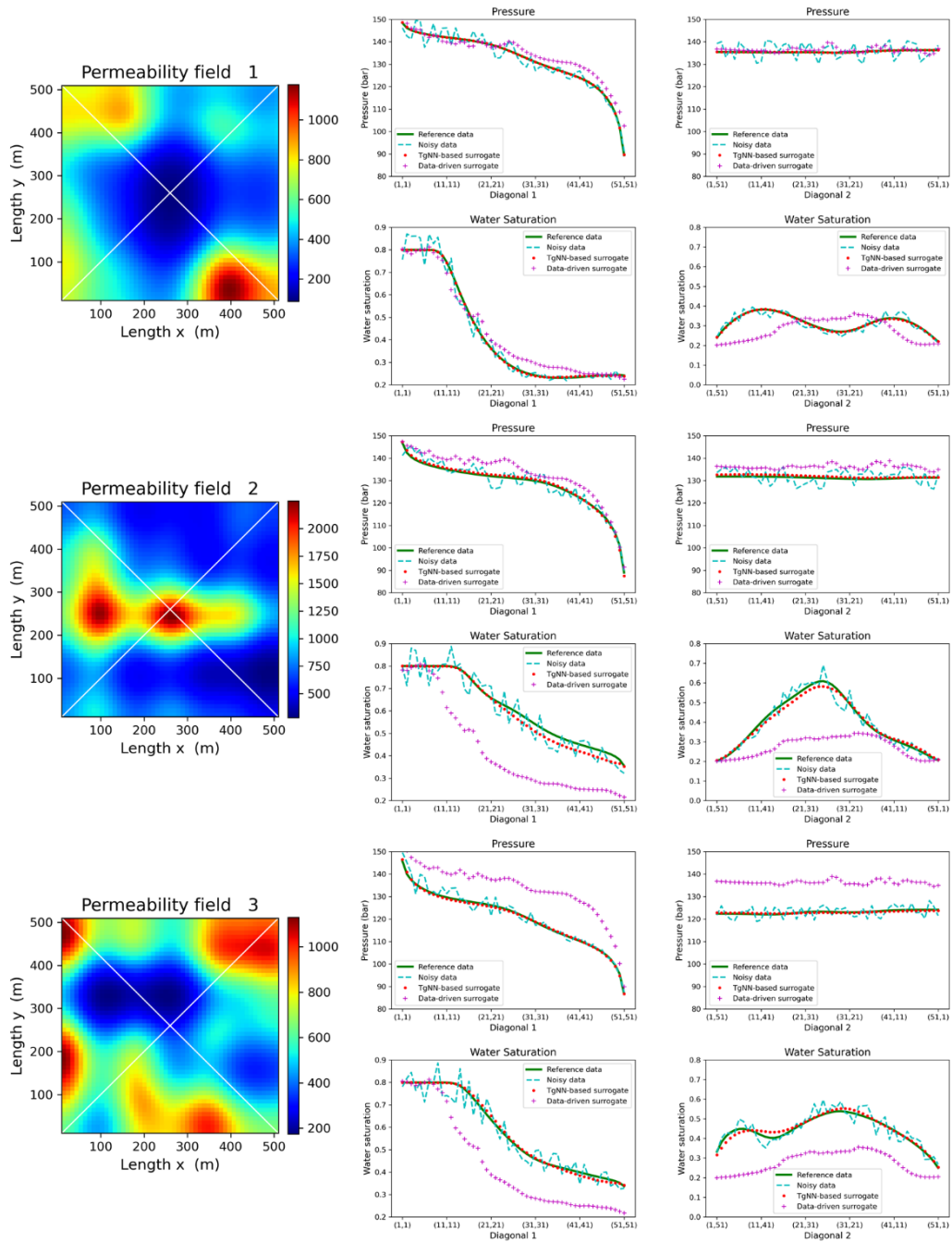

**Figure 9.** Predictions of pressure and saturation among the three stochastic permeability realizations on two diagonal lines at time-step 30.

**Figure 9** shows the predictions of pressure and saturation among the three stochastic permeability realizations on two diagonal lines at time-step 30. Two diagonal lines are defined to take into account the influence of source (injection well) and sink (production well) positions on the pressure and saturation fields. Diagonal line 1 is the line from the injection well to the production well and represents the drastic changes in the pressure and saturation fields, while diagonal line 2 is another line representing the overall changes in the pressure and saturation fields. It is obvious that the addition of noise causes fluctuations in the graph lines of the original reference. In the purely data-driven surrogate model, the predicted results fluctuate in the absence of other constraints. However, the predictions of the TgNN-based surrogate model are able to fit the reference data under the interference of noisy labeled data with the assistance of theoretical guidance. In the process of minimizing the composite loss function, the predictions of TgNN can obey known scientific principles and physical laws through the collocation points' calculation. Therefore, TgNN can reduce the effect of noise in the training stage and achieve more accurate predictions.

## 3.2 Increasing the heterogeneity of the permeability field

In section 3.1, the TgNN-based surrogate model has already worked well as an alternative for forward runs in UQ calculation. The complexity of permeability fields in reality, however, is much greater than that used in the previous cases. For this reason, in this section, more complex random fields are studied. To further investigate the generalization ability of the model, the heterogeneity of the permeability fields that we used is increased, i.e., the variance of the random field is extended, and the correlation length of the random field is changed.

### 3.2.1 Increasing the variance

The variance of permeability fields is changed from 0.5 to 1 in this case. In the

Karhunen-Loève expansion, changes in variance do not have an effect on the number of truncations in the formula, and thus the dimension of the input layer in the network remains unchanged. However, the increase in the complexity of the permeability fields may exacerbate the difficulty of training the neural networks. In order to address this, in the coupling part of the TgNN-based surrogate model, the neurons increase from 100 to 120 in each layer. 2,000,000 labeled data from 150 permeability fields are chosen for imposing constraints from data mismatch, and 6,000,000 collocation points from 500 permeability fields are chosen for embedding constraints from other scientific theories. The test set is 5,000 non-labeled permeability fields. **Figure 10** shows a comparative scatterplot of two random points taken from 200 permeability fields in the test set. It can be seen that the TgNN-based surrogate model predictions and the reference data are on a 45° line, indicating acceptable accuracy. In order to demonstrate a more comprehensive effect than the two points, a statistical histogram of the $R^2$ score error for the 500 non-labeled permeability fields in the test set is presented in **Figure 11**. As shown in the figure, the $R^2$ score of both pressure and water saturation is close to 1, which demonstrates the accuracy and robustness of the model.

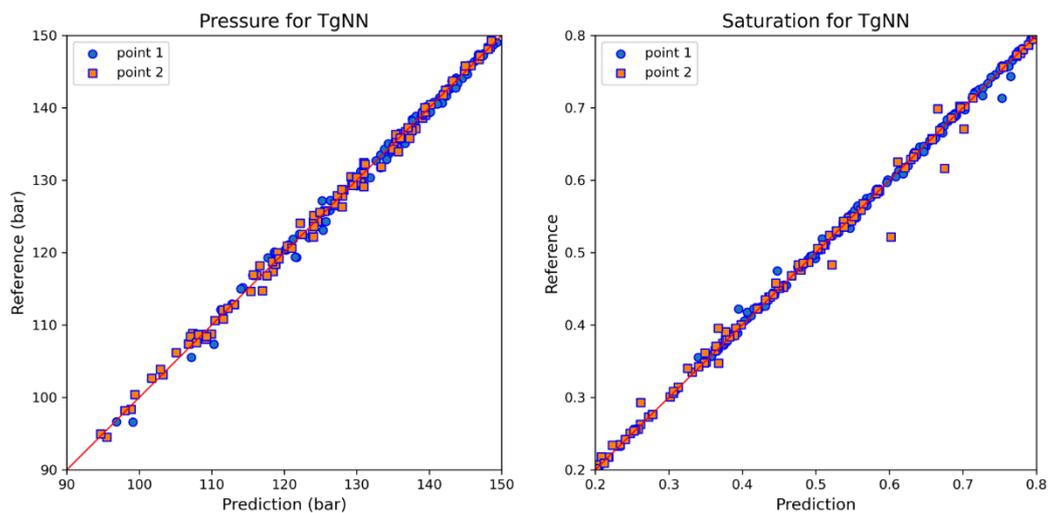

**Figure 10.** Comparison of two random points taken from the 200 non-labeled permeability fields in the test set with variance=1.0.

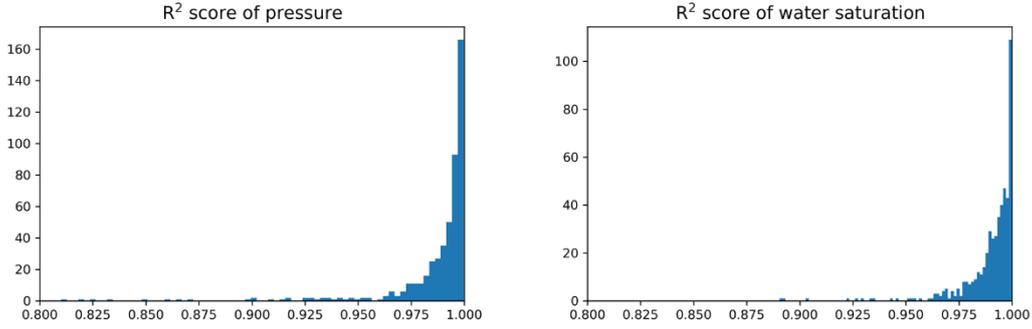

**Figure 11.** Statistical histogram of $R^2$ score for the 500 non-labeled permeability fields in the test set with variance=1.0.

### 3.2.2 Changing the correlation length

In section 3.1, the correlation length is 0.4 times the side length, which is 204 m in permeability fields. In this case, correlation lengths with different values are used to test the robustness of the network. Changes in the correlation length could affect the number of truncated terms in the formula of the Karhunen-Loève expansion. Therefore, in order to maintain 80% of the energy of the random permeability field, the number of dimensions should be changed accordingly. As a result, in the input layer of the TgNN-based surrogate model, the number of neurons varies with the correlation lengths.

The number of labeled permeability fields varies with the value of the correlation length. In order to maintain consistency, we uniformly selected 3,000,000 labeled data. The number of collocation points is 5,000,000 from 500 realizations. The test set is 5,000 non-labeled permeability fields. **Table 3** shows the predictive performance of the TgNN-based surrogate model under different correlation lengths, including the number of truncation terms that compose the permeability fields, the number of selected labeled realizations during training, and the mean relative $L_2$ error and mean $R^2$ score error of model prediction. **Figure 12** shows the comparison error of the mean and variance among 5,000 stochastic realizations at time-step 30 with different values of correlation length.

**Table 3.** Composition and prediction accuracy of the TgNN-based surrogate model under different correlation lengths.

| Correlation multiples | Correlation length | Number of truncated terms | Number of labeled realizations | $\overline{L_2}$ | $\overline{R^2}$ |
|---|---|---|---|---|---|
| 0.2 | 102 | 71 | 200 | 0.034474 | 0.94254 |
| 0.3 | 153 | 37 | 100 | 0.021935 | 0.983555 |
| 0.5 | 255 | 14 | 30 | 0.013133 | 0.995089 |

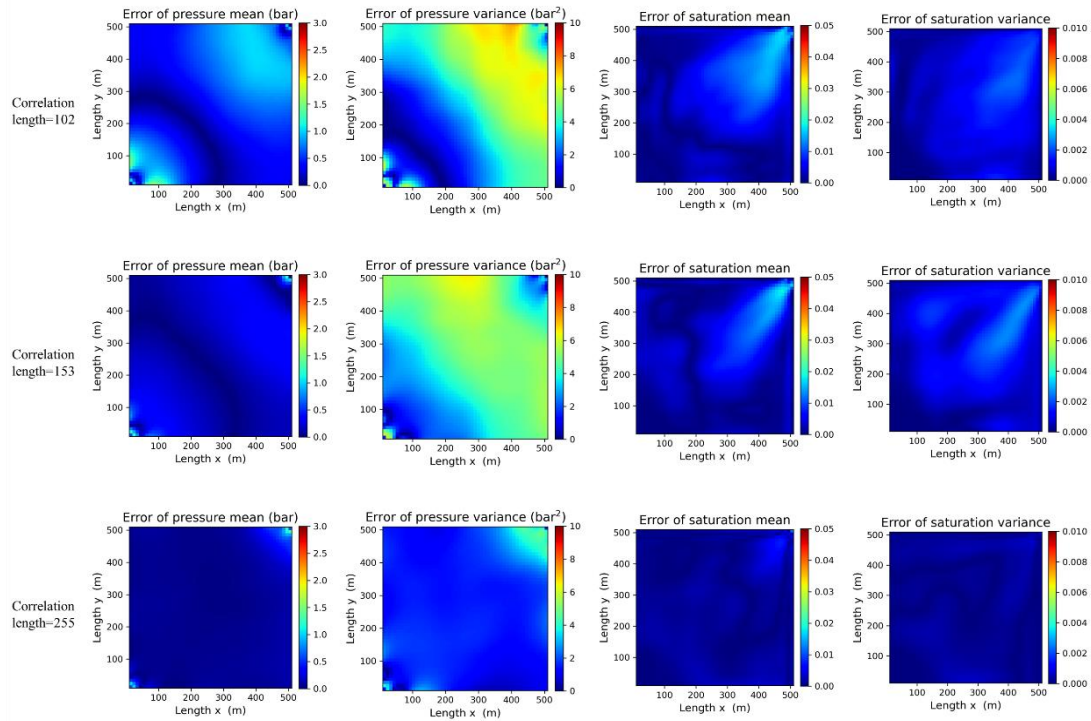

**Figure 12.** The Comparison error of the mean and variance among 5,000 stochastic realizations at time-step 30 with different values of correlation length.

It can be seen that changing the correlation length does affect the accuracy of the TgNN-based surrogate model. This is due to that the decrease in correlation length increases the number of requisite parameters to maintain the energy of stochastic permeability fields. This leads to an increase in the number of neurons in the input layer of the model which, in turn, affects the accuracy of the model. However, the TgNN-based surrogate model can still make satisfactory predictions for pressure and water saturation when faced with different configurations. The mean relative $L_2$ errors are close to 0, and the mean $R^2$ scores also approach 1. Comparing **Figure 4** with **Figure 12**, it can be seen that although the errors in pressure and saturation become larger with decreasing correlation length, they are still within acceptable limits.

## 3.3 Dynamically changing well conditions and variance of permeability fields

The previous cases are relatively stable in terms of condition constraints and stochastic realizations; however, many more complex examples arise in practical applications, such as varying well control conditions and permeability fields with varying degrees of heterogeneity. Keeping the network structure unchanged, we added two master control variables in this section, i.e., control conditions with different pressures and stochastic realizations with different variances.

**3.3.1 Dynamically changing well conditions**

In this case, we artificially add a varying well control pressure to the input layer to test whether the pressure can be accurately constrained into the network by the changed computational structure of the TgNN-based surrogate model. In the former section, the pressure in the formation at the initial moment is 100 bar, the pressure in the injection well is maintained to be 150 bar, and the pressure in the production well is maintained at 80 bar. In this section, however, the well control is randomly taken from 120 bar to 200 bar for injection wells and from 50 bar to 90 bar for production wells in each realization. Therefore, the input layer consists of $\{t_i$, $x_i$, $y_i$, $\xi_{i,1}$,

$\xi_{i,2}$, …, $\xi_{i,20}$, $P_{i,inj}$, $P_{i,pro}$ }, where $P_{i,inj}$ and $P_{i,pro}$ are the injection pressure and production pressure of $i$ th realization, respectively. We use 150 labeled permeability fields as the training set, and the number of neurons in the coupling part of the TgNN-based surrogate model increases from 100 to 120 in each layer, and that in the pressure part (i.e., $NN_P$ in **Figure 2**) of the model increases from 100 to 120 in each layer, respectively. 4,000,000 labeled data are chosen from the 150 labeled permeability fields, and 6,000,000 collocation points are chosen from the 500 permeability fields to constrain scientific theories.

**Figure 13** shows the matching performance of pressure and saturation among the two stochastic realizations on two diagonal lines at time-step 30 with varying well conditions. The estimated statistical moments and PDFs from 5,000 samples with different well conditions are shown in **Figures 14** and **15**, respectively. It can be seen that the TgNN-based surrogate model maintains satisfactory accuracy with the plot fitting the reference data. Although the pressure interval becomes larger due to the change in well control pressure, the TgNN can still provide accurate predictions and statistical quantity estimations for cases with dynamically changing well conditions.

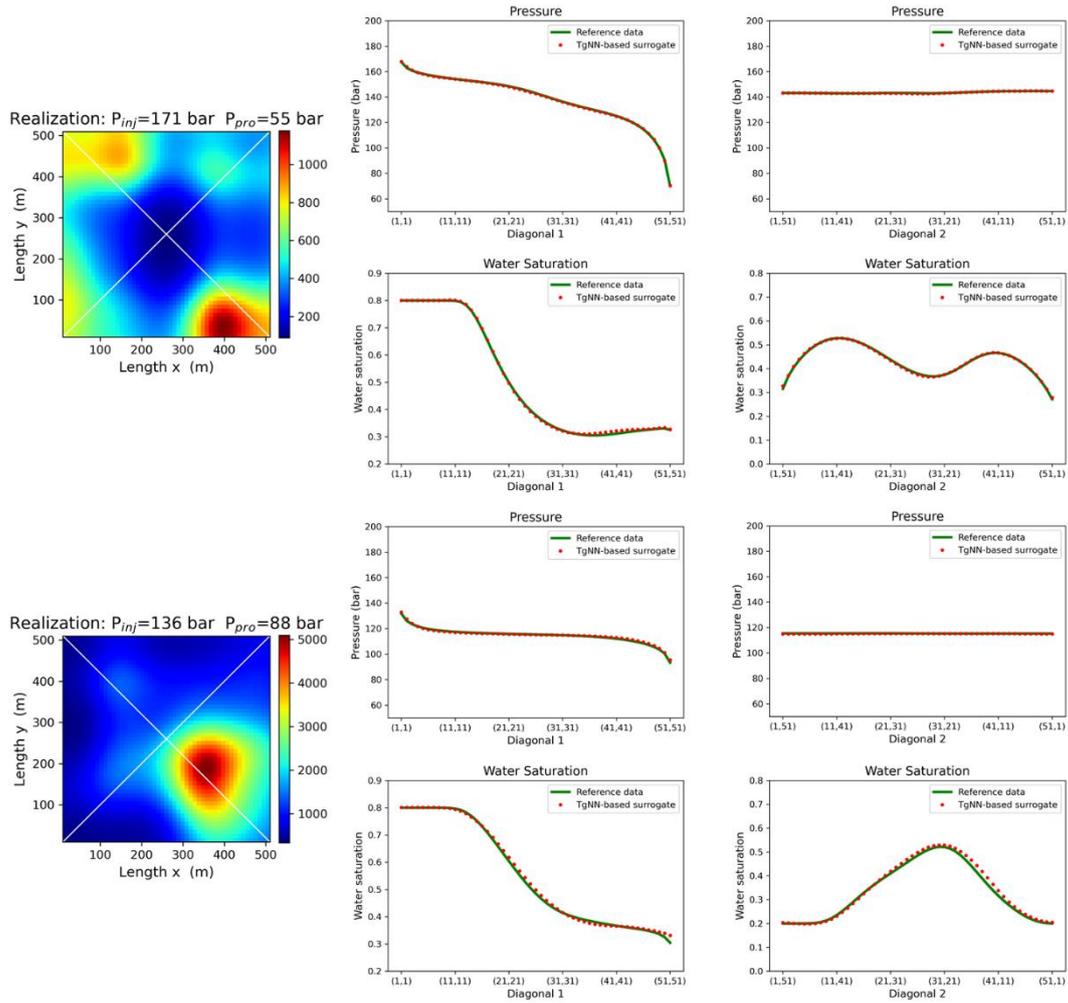

**Figure 13.** Predictions of pressure and saturation among the two stochastic realizations on two diagonal lines at time-step 30 with varying well conditions.

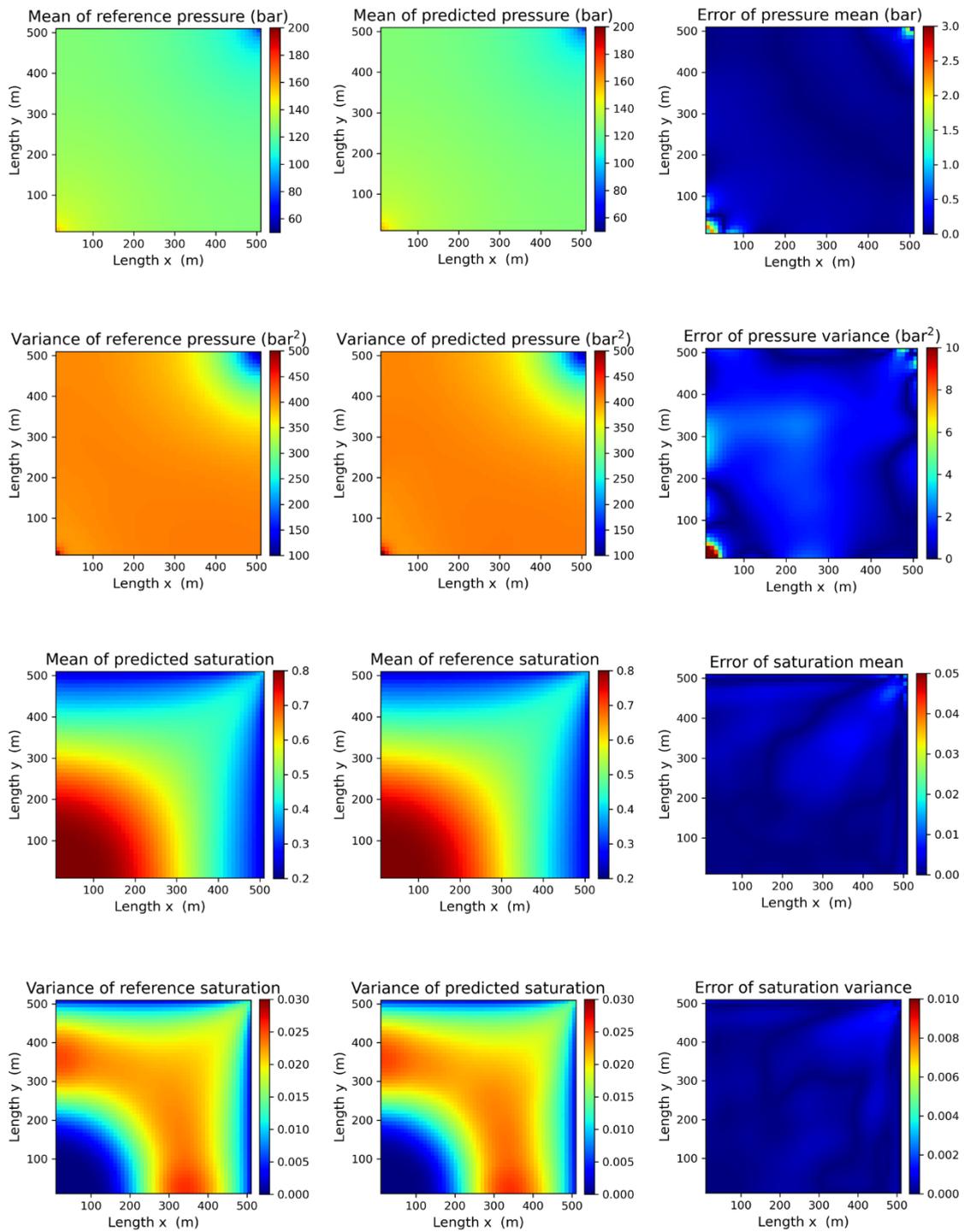

**Figure 14.** The mean and variance of pressure fields and saturation fields at time-step 30 between the TgNN-based surrogate model and the reference with varying well conditions.

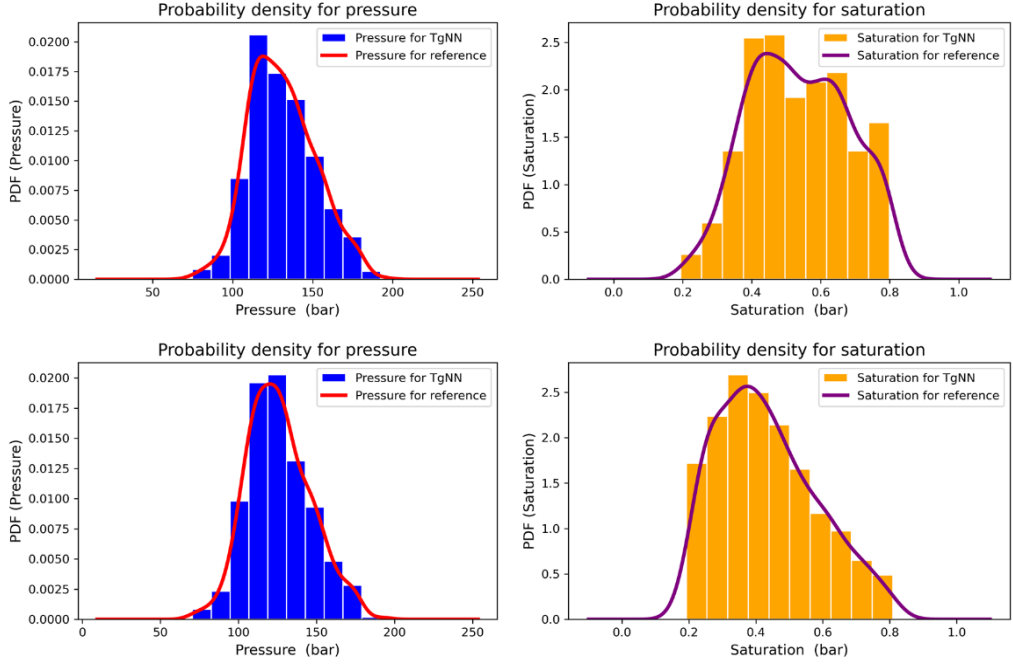

**Figure 15.** The estimated PDF of two fixed points (Point 1: x = 200 m, y = 200 m, t = 720 d; Point 2: x = 300 m, y = 400 m, t = 180 d) with varying well conditions.

**3.3.2 Dynamically changing variance of random fields**

In this case, we artificially add a varying variance of the permeability fields to the input layer to test whether this variance value can be accurately constrained into the permeability fields by the neural network, and then be constrained to the two-phase flow calculation through the TgNN surrogate operation structure. In the former sections, the variance of the permeability fields is a fixed value (0.5 in section 3.1 and 1 in section 3.2). In this case, the changing variance needs to be added to each collocation point of the training set. The variances of the stochastic fields are supposed to be uniformly distributed at (0.5-1), and the input layer consists of $\{t_i,\ x_i,\ y_i,\ \xi_{i,1},\ \xi_{i,2},\ \ldots,\ \xi_{i,20},\ \delta^2_i\}$, where $\delta^2_i$ is the variance of $i$ th realization. 250 labeled permeability fields are used as the training set, and the neurons in the coupling part of the TgNN-based surrogate model increase from 100 to 120 in each layer. 3,000,000 labeled data are chosen from the 250 labeled permeability fields, and 3,000,000 collocation points are chosen from 500 permeability fields to constrain scientific theories. **Figure 16** shows the matching performance of pressure and saturation among the two stochastic realizations on two diagonal lines with different variances. The estimated statistical moments and PDFs from 5,000 samples with dynamically changing variance of random fields are shown in **Figures 17** and **18**, respectively. It can be seen that the trained composite surrogate achieves superior performance in prediction for the case with dynamically changing variance of random fields.

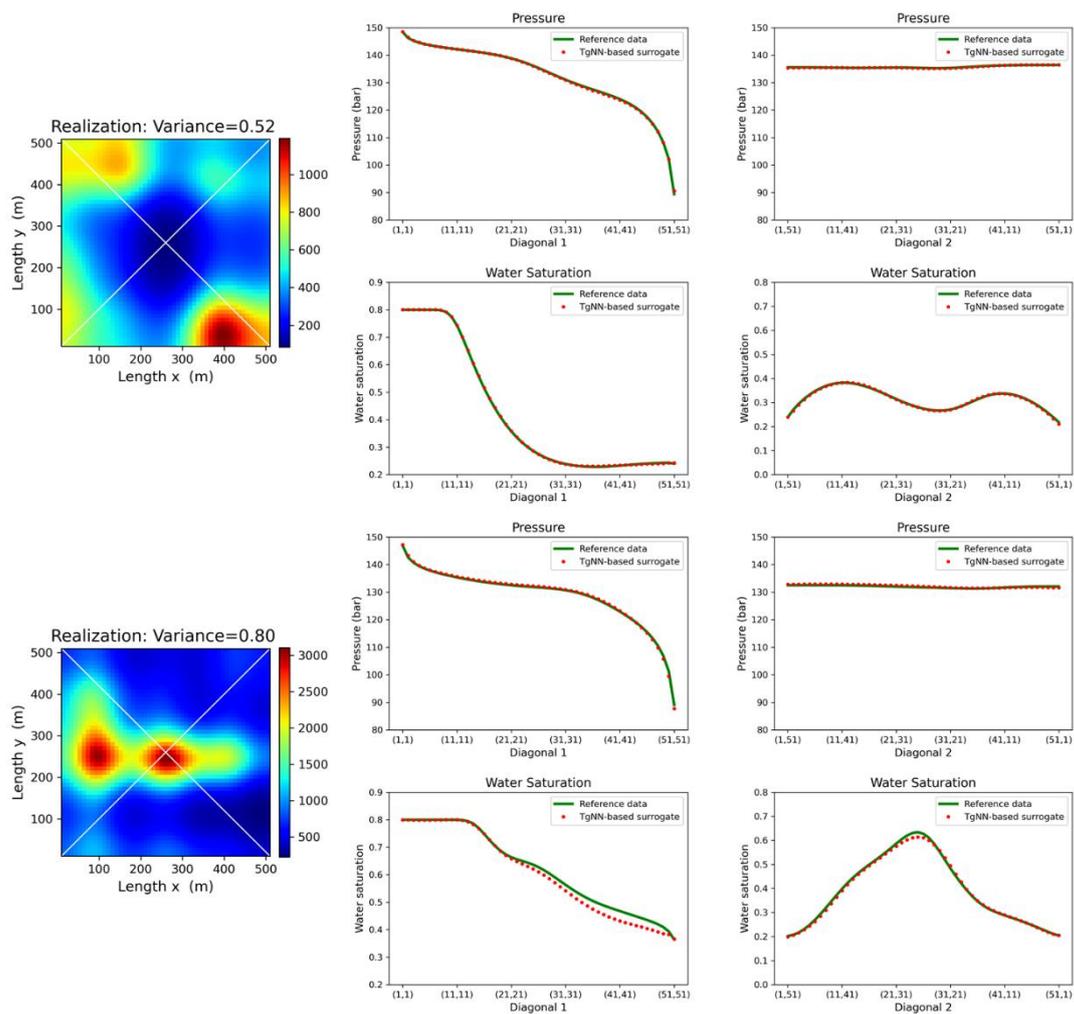

**Figure 16.** Predictions of pressure and saturation among the two stochastic realizations on two diagonal lines at time-step 30 with changing variance.

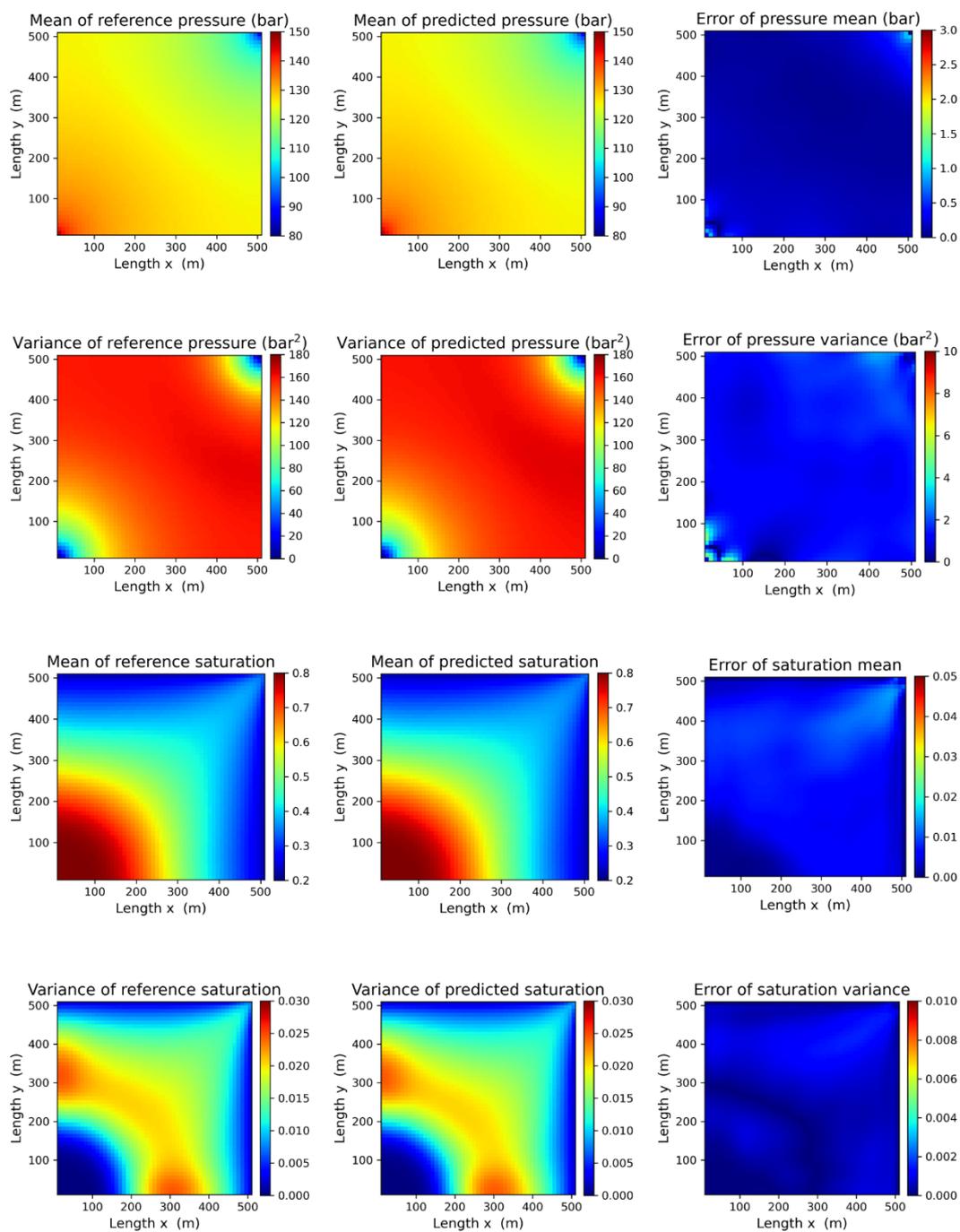

**Figure 17.** The mean and variance of pressure fields and saturation fields at time-step 30 between the TgNN-based surrogate model and the reference with changing variance.

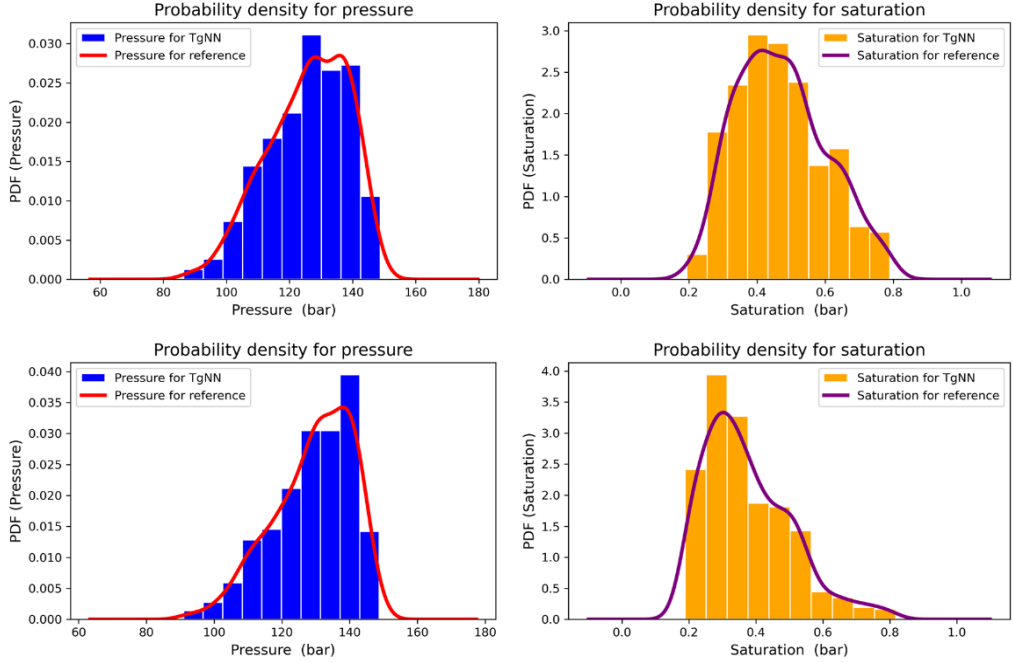

**Figure 18.** The estimated PDF of two fixed points (Point 1: x = 200 m, y = 200 m, t = 720 d; Point 2: x = 300 m, y = 400 m, t = 180 d) with changing variance.

## 4. Discussion and Conclusion

In this study, the TgNN-based surrogate model is proposed for two-phase flow in porous media, and the accuracy and robustness of the model are tested for uncertainty quantification. The input layer of the surrogate model consists of time-space coordinates, random field parameters, and other theoretical guidance. The output layers are pressure and water saturation, which not only comprise data mismatch, but also combine the residuals of the governing equations, initial conditions, boundary conditions, and expert knowledge into the loss function during training. In this paper, several network structures with different coupling forms are used to test the coupled two-phase flow equations, resulting in a new coupled neural network for oil-water phase flow problems, i.e., the TgNN-based surrogate model. The TgNN contains a coupled block and two independent blocks for approximating pressure and water saturation, respectively. Satisfactory results can be obtained by testing the accuracy and robustness of the TgNN model. By combining the accurate TgNN-based surrogate model with the Monte Carlo method, UQ tasks can be performed at a minimum cost to evaluate statistical quantities. The results show that the TgNN-based surrogate model is able to accurately capture the relationship between input and output well with high efficiency. Therefore, the trained TgNN can be used as a surrogate solution to the problem of predicting dynamic oil/water phase flow with new stochastic parameters.

Several scenarios are considered to test the accuracy and stability of this surrogate model in the UQ process, including the impact of labeled data and collocation points, the existence of data noise, increasing the variance and decreasing the correlation length of the permeability fields, changing well conditions, and changing the variance of the permeability fields. The increase of labeled data and collocation points will improve the accuracy of the surrogate model. Although a larger number of collocation points can improve the accuracy more significantly than that of labeled data, it also substantially increases computational cost. Therefore, in the process of training the surrogate model, we choose the strategy of taking as much valid labeled data as possible and a moderate number of collocation points. The addition of noise creates a fluctuation in the labeled data, which can further affect the data mismatch term during training. The TgNN-based surrogate model, however, can accurately capture the target features without relying entirely on the labeled data due to the constraint of theoretical guidance. In addition, the increased heterogeneity of the permeability fields in the input layer is also used to test the stability of the surrogate model. Irrespective of whether the variance is increased or the correlation length is decreased, the surrogate model can integrate more complex permeability fields well by modifying the neurons in the coupling layer of the network. Moreover, other theoretical guidance is added to the input layer, i.e., changing well conditions and changing the variance of permeability fields. The changing terms in the input layer are added as theoretical guidance via the governing equations, as well as the KL expansion used in parameterization for permeability fields.

The results demonstrate that the TgNN surrogate model is robust to problems with changing conditions of well control pressure and variance of stochastic fields.

Several unresolved issues and limitations exist in this work. On the one hand, the source-sink term in this work is only considered as one injection well and one production well, and does not involve scenarios with multiple wells and multiple forms of well placement. On the other hand, the realization of permeability fields in this article obeys the Gaussian distribution, and more complex non-Gaussian fields and more heterogeneous realizations need to be further researched. Since the fully connected layer requires a flattening of the original two-dimensional structure, which will destroy the original spatial structure and make it difficult for the fully connected layer to capture singularities with large structural changes (well points in this paper). However, the input of the image has a strong two-dimensional localized feature, which makes the variables (or pixels) in adjacent space or time highly correlated. Therefore, we will use CNN architectures in future work to construct surrogate models for the multiple wells problem. The permeability fields can be directly inputted in the form of images, and there is no need to implement automatic differentiation through the strong correlation of the permeability fields to the location information. Furthermore, the complex permeability fields can be well handled by manual finite differentiation, resulting in a two-dimensional two-phase flow surrogate model with non-Gaussian fields distribution of the permeability fields.